\documentstyle[nips99e,psfig,subfigure,amsmath,amssymb]{article}

\title{Sacrificial Learning in Nonlinear Perceptrons}

\author{Peixun Luo and K.~Y.~Michael Wong \\
Department of  Physics, 
Hong Kong University of Science and Technology\\
Clear Water Bay, Kowloon, Hong Kong \\
{\it \{physlpx, phkywong\}@ust.hk} \\
}

\begin{document}
\newcommand{\uD}{\mathrm{D}}
\newcommand{\avl}{\langle\!\langle}
\newcommand{\avr}{\rangle\!\rangle}

\maketitle

\begin{abstract}
Using the cavity method we consider the learning of noisy 
teacher-generated examples by a nonlinear student perceptron. 
For insufficient examples and weak weight decay, the activation 
distribution of the training examples exhibits a
gap for the more difficult examples. This illustrates 
that the outliers are sacrificed for the overall performance.
Simulation shows that the picture of the smooth energy landscape
cannot describe the gapped distributions well, implying that 
a rough energy landscape may complicate the learning process.
\end{abstract}

\section{Introduction}
The learning of noisy examples by a nonlinear perceptron is 
a {\it frustrating} process, in the sense that competing information 
extracted from the training examples needs to be processed
\cite{mpv}. Since learning usually involves minimizing 
a cost function, the learner often has to choose between 
interpolating the conflicting examples or sacrificing 
some in favor of satisfying others, so as to attain a minimum
overall cost. This kind of competition is especially marked in 
nonlinear perceptrons.

This sacrificial effect is an important issue. As we
shall see, it will lead to a gap in the activation 
distribution of the examples. The activations of the 
sacrificed examples are separated from those of
the satisfied ones by a wide margin, and different choices of 
partitioning the sacrificed and preferred  examples correspond to 
different local minima in the energy landscape. The appearance 
of a multiplicity of local minima leads to the roughening of the 
energy landscape, rendering the learning processes prone to being 
trapped before reaching the global minimum. 

The effect is best understood using 
the cavity approach \cite{epl, nips}. It uses a self-consistency argument
to consider what happens when a new example is added to the training set.
When the learner adopts a sacrificial learning strategy, learning the 
added example may lead to different choices of the sacrificed examples
in the background, causing a shift of the global energy minimum among a 
number of local minima. 

The cavity method yields identical macroscopic predictions with the replica
method \cite{mpv}. The assumption of a smooth energy landscape corresponds
to the replica symmetric ansatz in the replica approach. Instability
appears in the ansatz when the Almeida-Thouless condition
is violated \cite{mpv}, beyond which replica symmetry breaking solutions have
to be introduced, corresponding to a rough energy landscape. As we shall see,
the appearance of the gap is closely related to the Almeida-Thouless line.

In this paper, we analyze the learning of noisy examples by a nonlinear 
perceptron using the cavity method. We study the activation distribution of
the training examples and find parameter regimes with gaps 
in the distribution. Simulation results show that the assumption of a smooth
energy landscape works well when no gaps are present, but fails when gaps 
appear. 

\section{The cavity method} 

\subsection{Formulation}

The rule to be learned is generated by a teacher perceptron with $N$
weights $B_j$, $j=1,...,N$ and $\langle B_j^2\rangle=1$. 
The student perceptron, with $N$ weights $J_j$, $j=1,...,N$,
tries to model the teacher by learning from a set of $p$ 
examples. Each example, labeled $\mu$ with $\mu=1,...,p$, 
consists of an input vector ${\boldsymbol\xi}^\mu$ and the noisy output 
$O_\mu$ of the teacher. The input components $\xi_j^\mu$ are random variables,
with $\langle\xi_j^\mu\rangle=0$  
and $\langle\xi_j^\mu\xi_k^\nu\rangle=\delta_{jk}\delta_{\mu\nu}$.

The teacher output $O_\mu$ is a nonlinear activation function of the teacher 
activation $y_\mu\equiv{\boldsymbol B}\cdot{\boldsymbol\xi}^\mu/\sqrt{N}$,
corrupted by a Gaussian noise $\eta_\mu$ with $\langle\eta_\mu\rangle=0$ 
and $\langle\eta_\mu^2\rangle=1$. That is, $O_\mu\equiv f(y_\mu+T\eta_\mu)$, 
where $T$ is the noise level, and here we use the sigmoid function 
$f(x)=[1+e^{-x}]^{-1}$. Correspondingly, the student models the teacher 
outputs by $f_\mu\equiv f(x_\mu)$, where the student activation is 
$x_\mu\equiv{\boldsymbol J}\cdot{\boldsymbol\xi}^\mu/\sqrt{N}$.

Learning is attained by minimizing the energy function which consists of 
the errors of the student in reproducing the teacher's outputs for the 
training set, as well as the penalty term for excessive complexity.
Hence we use the energy function 
   \begin{equation}					\label{energy}
   E=\frac{1}{2}\sum_\mu(O_\mu-f_\mu)^2+\frac{{\bf\lambda}}{2}\sum_jJ_j^2,
   \end{equation} 
where $\lambda$ is the weight decay strength. Minimizing the energy 
function by gradient descent, the student reaches the equilibrium state 
   \begin{displaymath}
   J_j=\frac{1}{\lambda\sqrt{N}}\sum_\mu(O_\mu-f_\mu)f_\mu'\xi_j^\mu.
   \end{displaymath} 

\subsection{Adding an example}

If an example 0 is fed to the student, the activation 
   \begin{displaymath}
   t_0\equiv\frac{{\boldsymbol J}\cdot{\boldsymbol\xi}^0}{\sqrt{N}}
   \end{displaymath} 
is called cavity 
activation. Since the student ${\boldsymbol J}$ has no information about the 
example, the cavity activation is a Gaussian variable for random 
inputs $\xi_j^0$. Its mean, variance and covariance with the teacher activation
of example 0 are given by $\avl t_0\avr=0$, $\avl t_0^2\avr=q$ and 
$\avl t_0 y_0\avr=R$ respectively, where $\avl\cdot\avr$ 
denotes the ensemble average, and the parameters $q$ and $R$ are defined by 
   \begin{eqnarray}						  \label{qR}
   q=\langle J_j^2\rangle & \mathrm{and} & R=\langle J_j B_j\rangle. 
   \end{eqnarray}
Hence in the large $N$ limit, the cavity activation can be expressed as  
$t_0=R\,y_0+\sqrt{q-R^2}\,\zeta_0$,
where $\zeta_0$ is a Gaussian variable with mean 0 and variance 1.

Now compare the student ${\boldsymbol J}$ with another one which incorporates 
example 0 in the training set, denoted by ${\boldsymbol J}^0$. The generic 
student activation $x_0\equiv{\boldsymbol J}^0\cdot{\boldsymbol\xi}^0/
\sqrt{N}$ is no longer a Gaussian variable. Nevertheless, it is reasonable to 
expect that the difference between the students ${\boldsymbol J}$ 
and ${\boldsymbol J}^0$ is small. Following the perturbative analysis in \cite
{epl}, we conclude that $x_0$ is a well defined function of $t_0$, given by
   \begin{equation}
   t_0=x_0-\gamma(O_0-f_0)f_0',                                \label{act}     
   \end{equation} 
where $\gamma$ is the local susceptibility given by 
   \begin{equation}                                            \label{sus}
   1-\gamma\lambda=\alpha\left\langle 1-\frac{\partial{x_\mu}}
       {\partial{t_\mu}}\right\rangle_\mu; \quad \alpha=\frac{p}{N}. 
   \end{equation} 

For the nonlinear perceptron, it is possible that for 
sufficiently large $\gamma$, the generic activation $x_0$ is a multi-valued 
function of the cavity activation $t_0$. In this case we have to choose the 
one which minimizes the energy function in (\ref{energy}). The cavity method
shows that the energy increase on adding example 0 is 
   \begin{equation}					\label{change}
   \Delta E=\frac{1}{2}(O_0-f_0)^2+\frac{1}{2\gamma}(x_0-t_0)^2.
   \end{equation}
The first term is the primary change due to the added example, and the second 
term is due to the adjustment of the background examples. In the multi-valued
region one needs to compare those solutions whose values of $x_0$ are closer to
$t_0$ (therefore favorable in the background adjustment) with those whose 
outputs $f_0$ are closer to the teacher's outputs $O_0$ (therefore favorable 
in the primary cost). This competition leads to a discontinuity in the range 
of the activation $x_0$ when the cavity activation $t_0$ varies,
accompanied by the appearance of gaps in the activation distribution.

\subsection{Adding an input} 

Similarly, the cavity method can be used to analyze the changes when an 
input 0 is added to the rule. In this case, the teacher outputs are given 
by $O_\mu^0\equiv f(y_\mu+B_0\xi_0^\mu/\sqrt{N}+T\eta_\mu)$, where $y_\mu$
is the original teacher activation with
$N$ inputs. If the student has inputs 1 to $N$ only, the resultant student 
perceptron is given by 
$J_j=\sum_\mu(O_\mu^0-f_\mu)f_\mu'\xi_j^\mu /(\lambda\sqrt{N})$.

We may construct the weight 0 for the student perceptron using the same 
prescription, namely 
   \begin{displaymath}
   Z_0=\frac{1}{\lambda\sqrt{N}}\sum_\mu(O_\mu^0-f_\mu)f_\mu'\xi_0^\mu.
   \end{displaymath} 
However, this is not the generic weight since in the activation functions 
$f_\mu$, the arguments $x_\mu$ do not contain the input 0; nor is the 
information fed from input 0 ever being utilized in the learning of $x_\mu$.
Hence $Z_0$ is called the cavity weight.

Now compare the student with another one with inputs 0 to $N$, and which 
incorporates input 0 in all the training examples. Its weights are denoted 
by $J_j^0$ for $j=1,...,N$ and $J_0$ for input 0. $J_0$ is different from 
$Z_0$. Nevertheless, it is also reasonable to expect that the difference 
between the students $J_j$ and $J_j^0$ is small. Using the perturbative 
analysis, we conclude that $J_0$ is a well defined function of $Z_0$, given 
by 
   \begin{equation}					 \label{weight}
   J_0=\gamma\lambda Z_0.
   \end{equation}
The cavity weight distribution $P(Z_0|B_0)$ is a Gaussian with mean 
and variance given by 
   \begin{equation}					\label{zmean}	
   \avl Z_0\avr=\frac{\alpha}{\lambda}\left\langle\frac{O_\mu'f_\mu'}{1+\gamma
    [f_\mu^{\prime 2}-(O_\mu-f_\mu)f_\mu'']}\right\rangle_\mu B_0,
   \end{equation}
   \begin{equation}					\label{zvar}
   \avl Z_0^2\avr-\avl Z_0\avr^2=\frac{\alpha}{\lambda^2}\langle(O_\mu-f_\mu)^2
    f_\mu^{\prime 2}\rangle_\mu.
   \end{equation}

\subsection{Macroscopic parameters}

Making use of the relation (\ref{weight}), 
we can obtain the self-consistent equations for the macroscopic
parameters $\gamma$, $R$ and $q$ in (\ref{sus}) and (\ref{qR}),
   \begin{eqnarray} 
   1-\gamma\lambda&=&\alpha\gamma\int\!\!\!\!\int
        \frac{f'(x)^{2}-[f(\sqrt{1+T^2}u)-f(x)]f''(x)}{1+\gamma
   \{f'(x)^{2}-[f(\sqrt{1+T^2}u)-f(x)]f''(x)\}}{\rm D} {\it u}{\rm D} 
       {\it v},        \\
   R&=&\alpha\gamma\int\!\!\!\!\int\frac{f'(\sqrt{1+T^2}u)f'(x)}
        {1+\gamma\{f'(x)^{2}-[f(\sqrt{1+T^2}u)-f(x)]f''(x)\}}{\rm D} 
       {\it u}{\rm D} {\it v},      \\
   q-R^2&=&\alpha\gamma^2\int\!\!\!\!\int
        [f(\sqrt{1+T^2}u)-f(x)]^2 f'(x)^2{\rm D} 
       {\it u}{\rm D} {\it v},
  \end{eqnarray}
where D$u\equiv$d$u\exp[-u^2/2]/\sqrt{2\pi}$ 
and D$v\equiv$d$v\exp[-v^2/2]/\sqrt{2\pi}$ 
are two independent Gaussian measures. Both the noise-corrupted teacher 
activation and the cavity activation are Gaussian distributed and determined 
from $u$ and $v$ via 
   \begin{equation}
   y+T\eta=\sqrt{1+T^2}\,u  \quad {\mathrm and} \quad 
        t=\frac{R}{\sqrt{1+T^2}}u+\sqrt{q-\frac{R^2}{1+T^2}}\,v,
   \end{equation}
and the generic activation $x$ is given by the solution of 
   \begin{equation}
   t=x-\gamma\,[f(\sqrt{1+T^2}u)-f(x)]f'(x).
   \end{equation}

The progress of learning is monitored by the training error $\epsilon_t$ and
the generalization error $\epsilon_g$, which are respectively the root mean 
square errors for a training example and an arbitrary example,
  \begin{eqnarray}   
  \varepsilon_t^2&=&\int\!\!\!\!\int[f(\sqrt{1+T^2}\,u)-f(x)]^2{\rm D} 
       {\it u}{\rm D} {\it v},	 \\
  \varepsilon_g^2&=&\int\!\!\!\!\int\left[f(\sqrt{1+T^2}\,u)-f
  \left(\frac{R}{\sqrt{1+T^2}}u+\sqrt{q-\frac{R^2}{1+T^2}}\,v\right)\right]^2
   {\rm D}{\it u}{\rm D}{\it v}. 
  \end{eqnarray}

\subsection{Stability condition}

The validity of the perturbation approach can be checked by considering
the stability condition of the equilibrium state. When example 
0 is added, the amplitude of the change in the student vector is given by 
   \begin{equation}
   \sum_j(J_j^0-J_j)^2=\frac{(x_0-t_0)^2}{1-\alpha\left\langle\left(1-\frac
   {\partial x_\mu}{\partial t_\mu}\right)^2\right\rangle_\mu}.
   \end{equation}
Hence the stability condition is   
   \begin{equation}						\label{stab}
   \alpha\left\langle\left(1-\frac{\partial x_\mu}{\partial t_\mu}
     \right)^2\right\rangle_\mu<1. 
   \end{equation} 
This is identical to the stability condition of the replica-symmetric 
ansatz in the replica approach, the so-called Almeida-Thouless (AT) 
condition. In particular, we note that when a gap is present in the activation
distribution, $x_\mu$ is a discontinuous function of $t_\mu$ and the system 
becomes unstable.

\section{The activation distribution}

Gaps in the activation distribution appear for values of local susceptibility 
$\gamma>16.6$, when a single value of the cavity activation in (\ref{act})
may map onto multiple values of the generic activation, corresponding to 
different energy minima. When the energy minimum 
favors the generic activation to take a value closer to the teacher activation
than the cavity activation, the example is satisfied. Otherwise, when the 
generic activation is closer to the cavity activation, the example is 
sacrificed. 

As shown in Fig.~1(a), sacrificial learning first occurs at the extreme values 
of the teacher output $O$. This is because in nonlinear perceptrons, 
changes in the student activation around these extreme values of $O$ do not 
result in significant changes in the training error of an example, and if the 
cavity activation is very different from the teacher's, it is more economical 
to keep the student activation close to the cavity activation, so that the 
background  adjustment remains small. Hence sacrificial learning is a unique 
consequence of the nonlinearity of the perceptron output. In contrast, no 
sacrificial learning is present in linear perceptrons, even when perfect 
learning is impossible \cite{Bos}. 

\begin{figure} [hbt]
\centering
\leftline{\psfig{figure=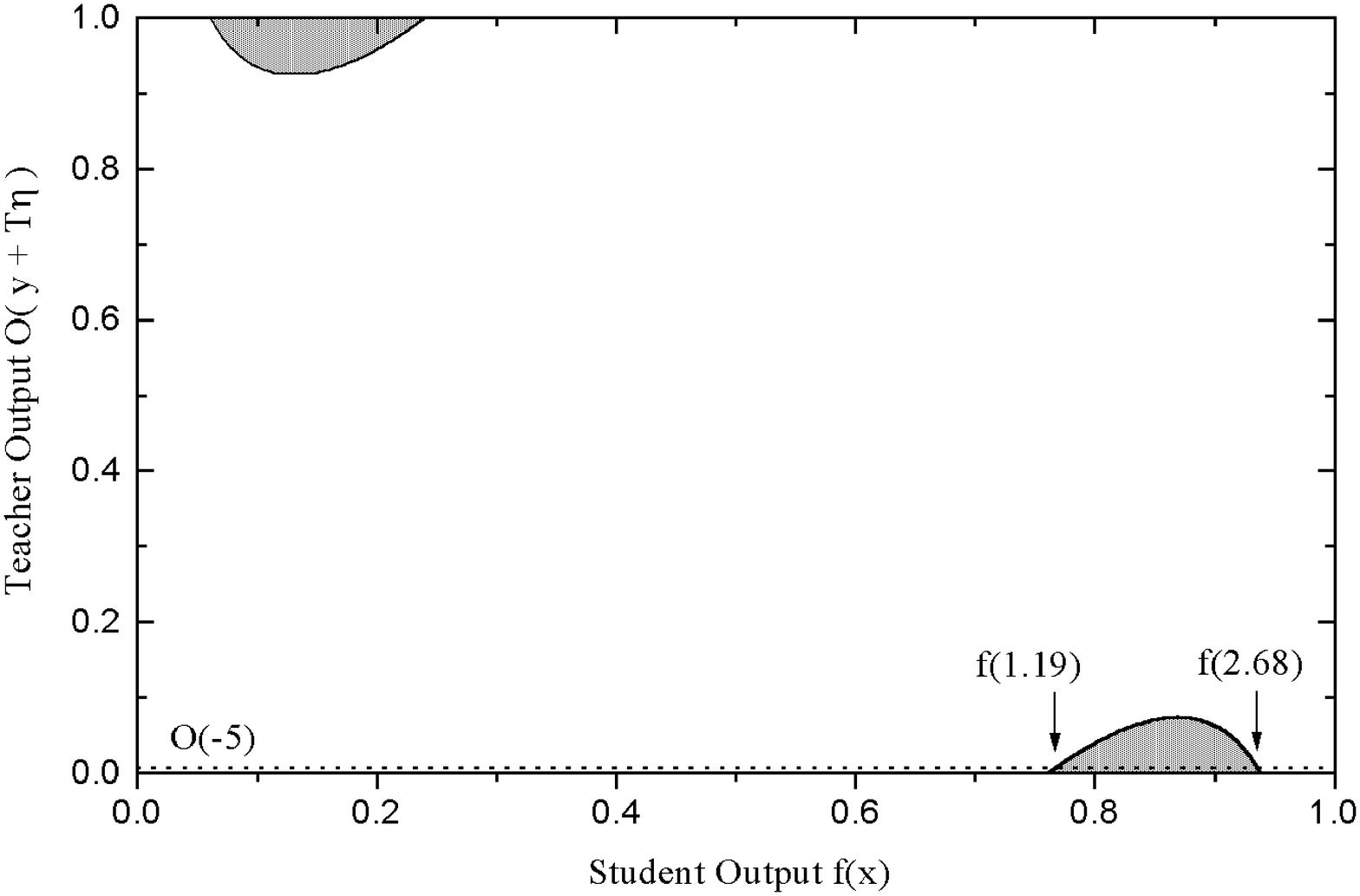,height=6.2cm,width=6.2cm}}
\vspace*{-6.2cm}
\leftline{\hspace*{6.2cm}
{\psfig{figure=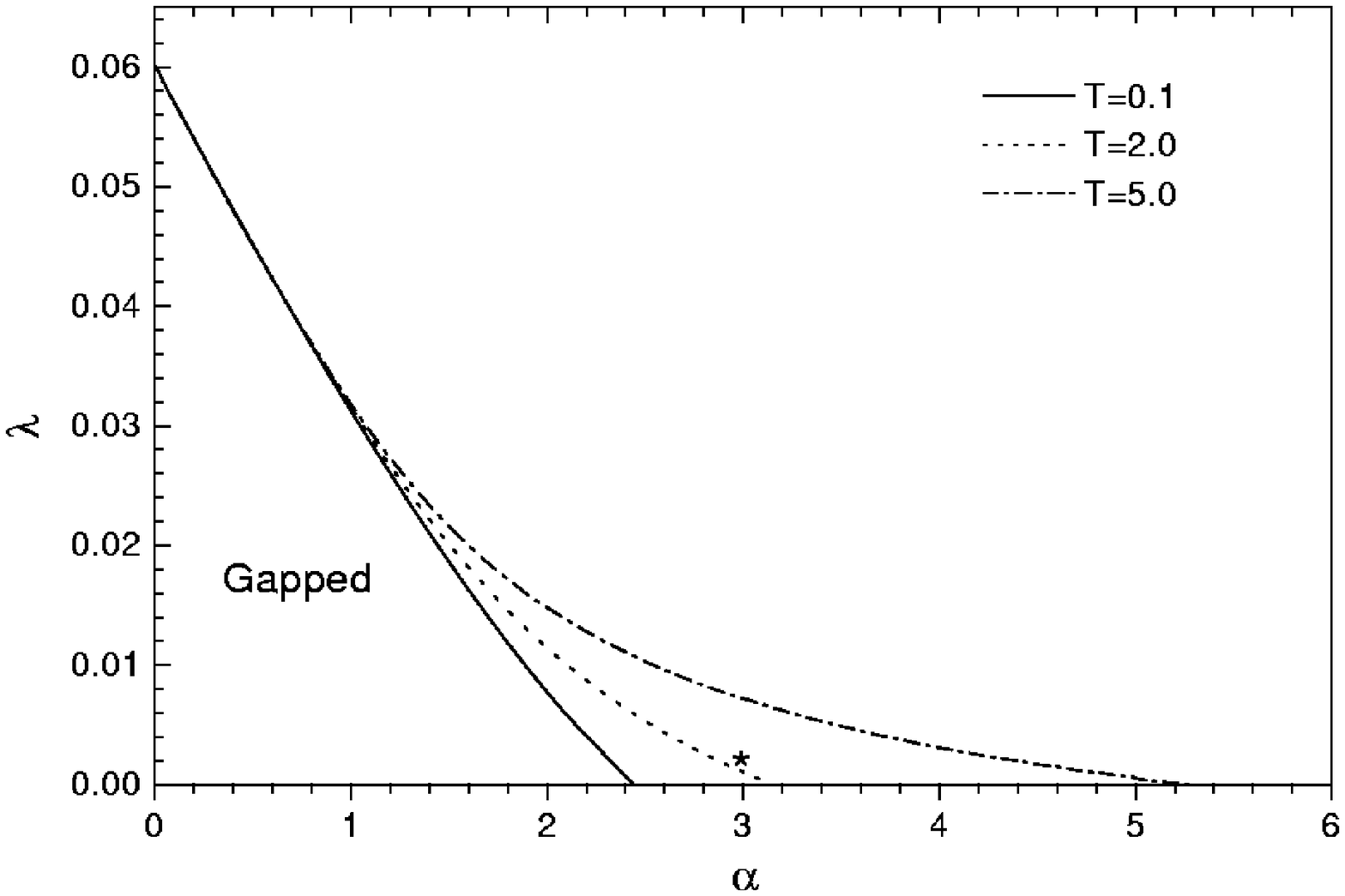,height=6.2cm,width=6.2cm}}}
\caption{
(a) The occurrence of sacrificial learning when $\gamma=18.6$. No values of the
student output exist in the shaded region. Here $\alpha=3$, $T=2$ and 
$\lambda=0.002$. Dotted line: $y+T\eta=-5$, corresponding to the 
distribution in Fig.~2(b).
(b) Regions of the existence of gapped activation distribution for different 
noise levels. The point $\alpha=3$, $\lambda=0.002$ is denoted by a star.
}
\end{figure}

In Fig.~1(a), no values of the student output in the shaded region exist. 
For $O<0.078$, student activations to the left of the shaded region 
correspond to the satisfied examples, whereas those to the right correspond 
to the sacrificed ones. For intermediate values of $O$, the competitive 
effects are less, and there are no gaps developed. 

Figure 1(b) shows the regions for the existence of gapped activation 
distributions. They are closely related to the unstable regions which violate 
the condition (\ref{stab}). The gapped regions lie inside the unstable regions,
since the development of a gap is already sufficient to cause an 
uncontrollable change when a new example is added. However, provided that 
$\alpha$ is not too small, the boundaries of the gapped and unstable regions 
are very close to each other. 

Figure 1(b) shows that frustration is serious when the training set size 
is small, leading to gapped activation distributions. When the training 
examples are sufficient, the underlying rule can be extracted with confidence,
thereby restoring the continuous distribution. Furthermore, increasing the 
data noise broadens the gapped region. Indeed, noisy data introduces competing
information to be learned by the student, and hence increases the degree of 
frustration. On the other hand, the gapped region narrows with increasing 
weight decay strength. Arguably, weight decay restricts the flexibility in the 
weight space, thus reducing the tendency for multiple minima.

\begin{figure} [hbt]
\centering
\leftline{\psfig{figure=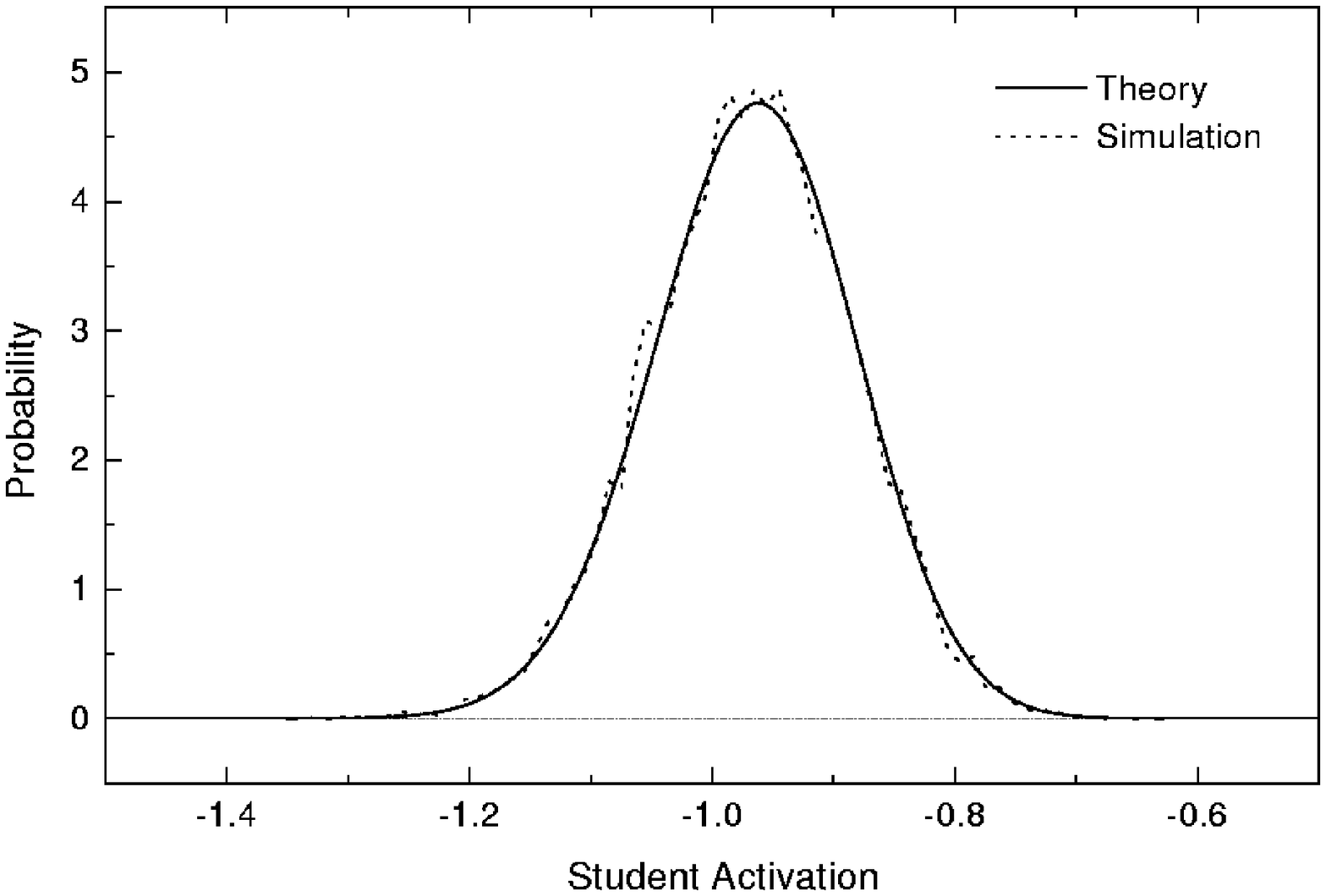,height=6.0cm,width=6.2cm}}
\vspace*{-6.0cm}
\leftline{\hspace*{6.2cm}
{\psfig{figure=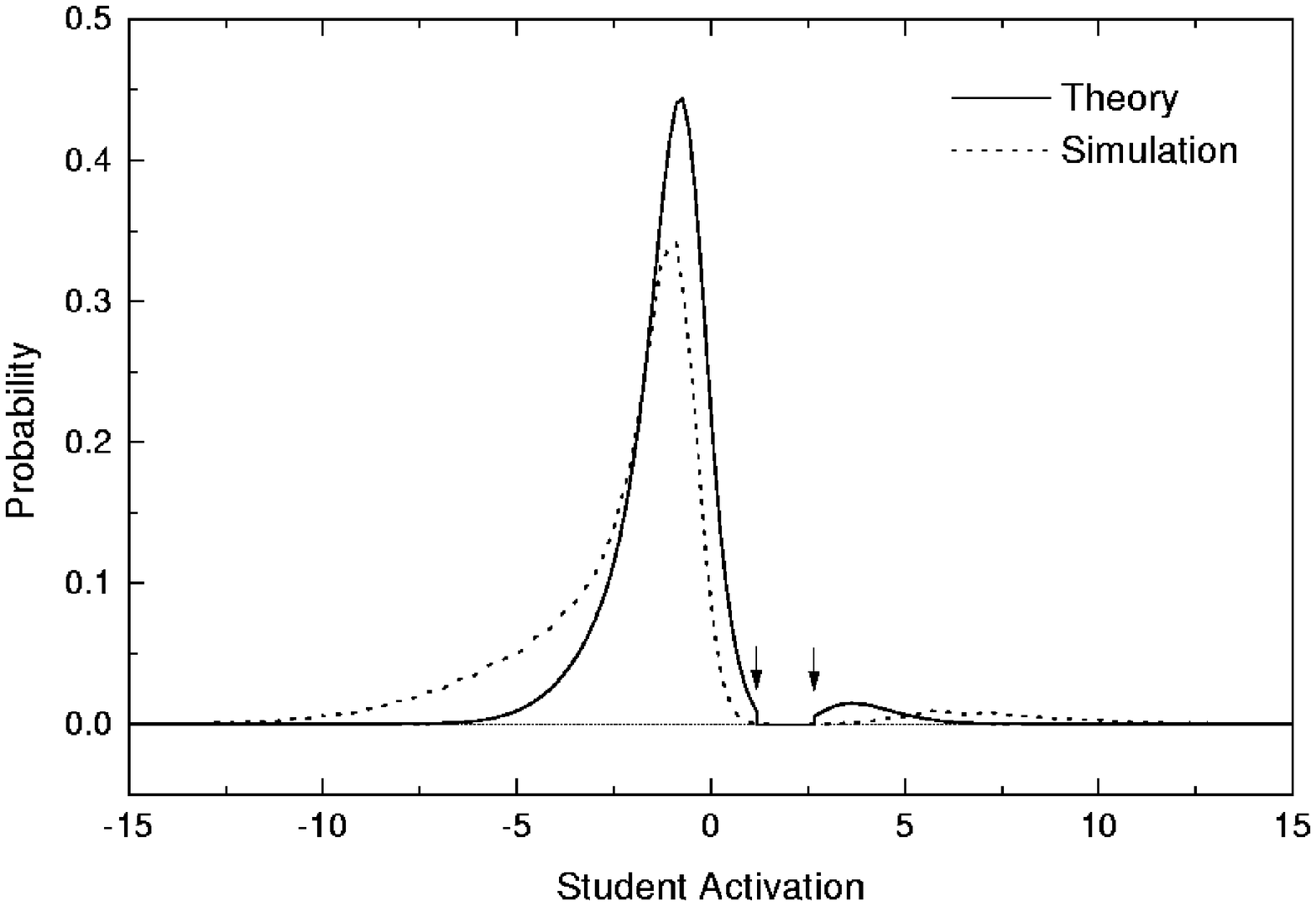,height=6.0cm,width=6.2cm}}}
\caption{
Student activation distribution at $\alpha=3$, and $\lambda=0.002$
(denoted by a star in Fig.~1(b)).
(a) $T=0.1$ and  $y+T\eta=-1$;
(b) $T=5$ and $y+T\eta=-5$. 
Note that in (b) the gap from the simulation is broader than the theoretical 
range $[1.19, 2.68]$ (arrows, see Fig.~1(a)).
}
\end{figure}

Figure 2(a) shows a typical activation distribution 
in the region of continuous distribution, where $\gamma=11.1$ and the 
stability condition (\ref{stab}) is fulfilled. Comparing with the 
simulation result, we see that the assumption of a smooth energy landscape 
used in the present work is valid in this region. The theoretical and 
simulational results of $\epsilon_t$ and $\epsilon_g$ also agree. In contrast, 
the gapped distributions in Fig.~2(b) show that the 
assumption does not well describe the simulation result when $\gamma=18.6$ 
and the stability condition (\ref{stab}) is violated. There are prominent
differences of $\epsilon_t$ and $\epsilon_g$ between theoretical 
and simulational results. To improve the agreement, a rough energy 
landscape as discussed in \cite{nips} must be introduced.

\section{Conclusion}

We have demonstrated the existence of band gaps in the activation distribution,
and attributed them to frustrations arising from the competition of conflicting
information inherent in noisy data, and the nonlinearity of the student 
perceptron. Activations corresponding to sacrificed or satisfied examples 
during learning are seperated by band gaps. The existence of band gaps 
necessitates the picture of a rough energy landscape. In the picture of the 
replica approach, it corresponds to the replica symmetry breaking ansatz 
beyond the Almeida-Thouless line. 

We remark that the sacrificial effects are common in many other cases, 
such as multilayer perceptrons \cite{nips} and weight pruning \cite{tanc}.
It may also exist in Support Vector Machines (SVM) when examples are noisy and 
insufficient \cite{Vapnik}. They may create local minima which complicate 
the convergence of learning processes. Hence it is an issue that should be 
considered both theoretically and practically.

\subsubsection*{Acknowledgments}
This work is supported by the Research Grant Council of Hong Kong (HKUST6157
/99P).

\end{document}